\documentclass[reprint,aps,prl,superscriptaddress]{revtex4-1}

\pdfoutput=1

\usepackage{graphicx}
\usepackage[dvipsnames]{xcolor}
\usepackage{calc} 

\usepackage{amsmath}
\usepackage{amssymb}
\usepackage{bm}

\usepackage{braket}
\usepackage{units}
\usepackage{ mathrsfs}

\usepackage{eurosym}
\usepackage{cancel}
\usepackage{upgreek}

\usepackage{url}
\usepackage[colorlinks=true,
        linkcolor=black,
        citecolor=black,
        filecolor=black,
        urlcolor=black,
        bookmarks=false,
        bookmarksopen=false,
        bookmarksopenlevel=3,
        plainpages=false,
        pdfpagelabels=true,hyperfootnotes,
        pdfauthor={J. A. Mlynek},
        pdftitle={Observation of Dicke Superradiance for Two Artificial Atoms in a Cavity with High Decay Rate}
        ]{hyperref}

\newcommand{\chd}[1]{{\color{black}{#1}}}

\begin{document}

\title{Observation of Dicke Superradiance\\ for Two Artificial Atoms in a Cavity with High Decay Rate}


\author{J.~A.~Mlynek}
\author{A.~A.~Abdumalikov Jr}
\author{C.~Eichler}
\author{A.~Wallraff}
\affiliation{Department of Physics, ETH Z\"urich, CH-8093, Z\"urich, Switzerland.}
\date{\today}

\maketitle

\textbf{
An individual excited two level system decays to its ground state by emitting a single photon in a process known as spontaneous emission \cite{Haroche2006}. In accordance with quantum theory the probability of detecting the emitted photon decreases exponentially with the time passed since the excitation of the two level system. In 1954 Dicke first considered the more subtle situation in which two emitters decay in close proximity to each other \cite{Dicke1954}. He argued that the emission dynamics of a single two level system is altered by the presence of a second one, even if it is in its ground state.
Here, we present a close to ideal realization of Dicke's original two-spin Gedankenexperiment, using a system of two individually controllable superconducting qubits weakly coupled to a microwave cavity with a fast decay rate. The two-emitter case of superradiance is explicitly demonstrated both in time-resolved measurements of the emitted power and by fully reconstructing the density matrix of the emitted field in the photon number basis.}

\begin{figure}[t]
  \centering
  \includegraphics[width=0.5\textwidth]{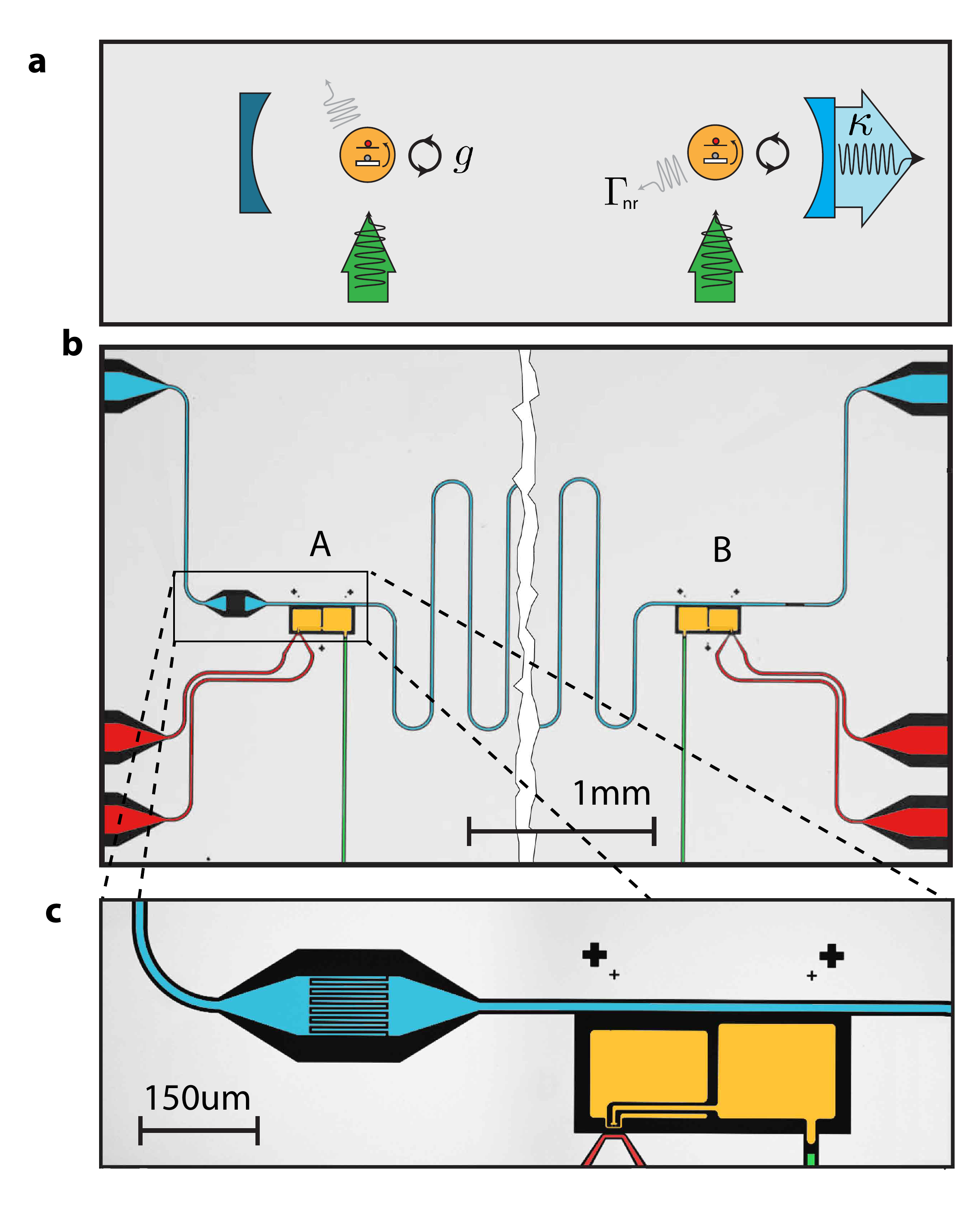}
  \caption{\textbf{Schematic and Sample.} \textbf{a}, Optical equivalent of the setup. Two two-level systems (yellow) are coupled with identical rate $g$ and intrinsic decoherence rate $\Gamma_{nr}$ to a cavity mode with photon decay rate $\kappa$ (blue). The two-level systems are excited by radiation applied orthogonal to the cavity mode (green).  \textbf{b,c}, Optical microscope false color image of the sample with two qubits (A,B) (yellow) capacitively coupled to an asymmetric waveguide resonator (blue). Each qubit is equipped with a local charge gate (green) and a flux bias line (red) to create initial states and tune transition frequencies independently.}
  \label{fig:sample}
\end{figure}
Since 1954, enhanced superradiant decay has been observed in many different physical systems \cite{Skribanowitz1973,gross1976,raimond1982,Scheibner2007,Rohlsberger2010}. The obtained results are consistent with Dicke's prediction that the emitted power of large ensembles depends on the square of the density of the emitters rather than showing a linear dependence. However, for large numbers of atoms or atom-like systems a direct observation of superradiance may be hindered by numerous impeding effects, such as nonlinear propagation and diffraction which occur in dense ensembles \cite{Gross1982}.

Striving to realize ideal conditions for its observation, a number of experiments were designed to explore the microscopic regime of superradiance by employing a small number of two-level emitters \cite{DeVoe1996,Eschner2001a, Filipp2011a,vanLoo2013}. In particular experiments involving two trapped ions were able to show that their collective decay rate varied by a few percent depending on their separation \cite{DeVoe1996,Eschner2001a}. These experiments presented clear evidence of an enhanced decay, but were unable to resolve the dynamics by directly measuring the intensity of the emitted radiation as a function of time. Although the ions could be driven directly into either sub- or superradiant states, arbitrary initial states could not be directly prepared. In addition, the observed superradiant decay did not dominate over other decay mechanisms, because of the too large distance $R$ between the emitters exceeding the wavelength $\lambda$ of the emitted radiation.

\begin{figure}[t]
  \centering
  \includegraphics[width=0.5\textwidth]{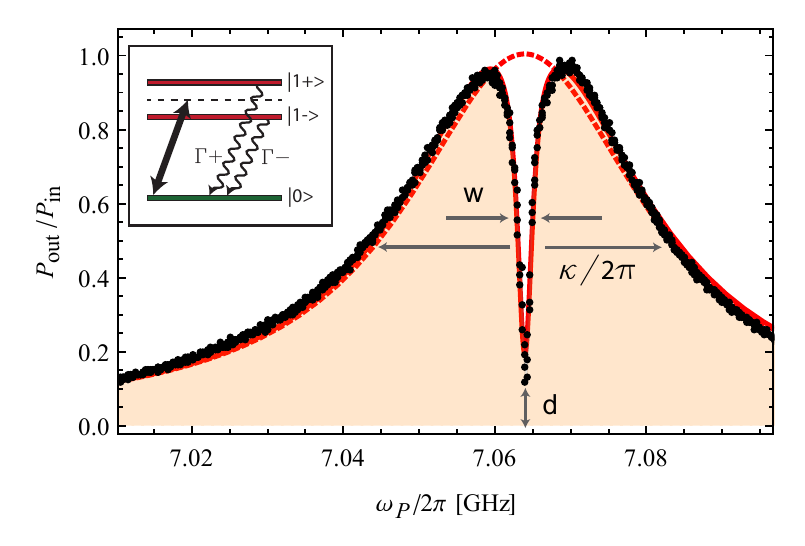}
  \caption{\textbf{Resonator Transmission Spectrum.} Resonator transmittance $P_\textrm{out}/P_\textrm{in}$
  extracted from a narrow band homodyne measurement with qubit A tuned into resonance with the large decay rate cavity (black dots) compared to an analytical model (solid red line), see text for details. The Lorentzian spectrum of the resonator with the two-level systems largely detuned is indicated by a dashed red line. The excited state doublet $|1+\rangle,|1-\rangle$ with two decay paths $\Gamma_{\pm}$ to the ground state $|0\rangle$  is shown in the inset.}
  \label{fig:transspec}
\end{figure}

\begin{figure*}[t!]
  \centering
  \includegraphics[width=1.0\textwidth]{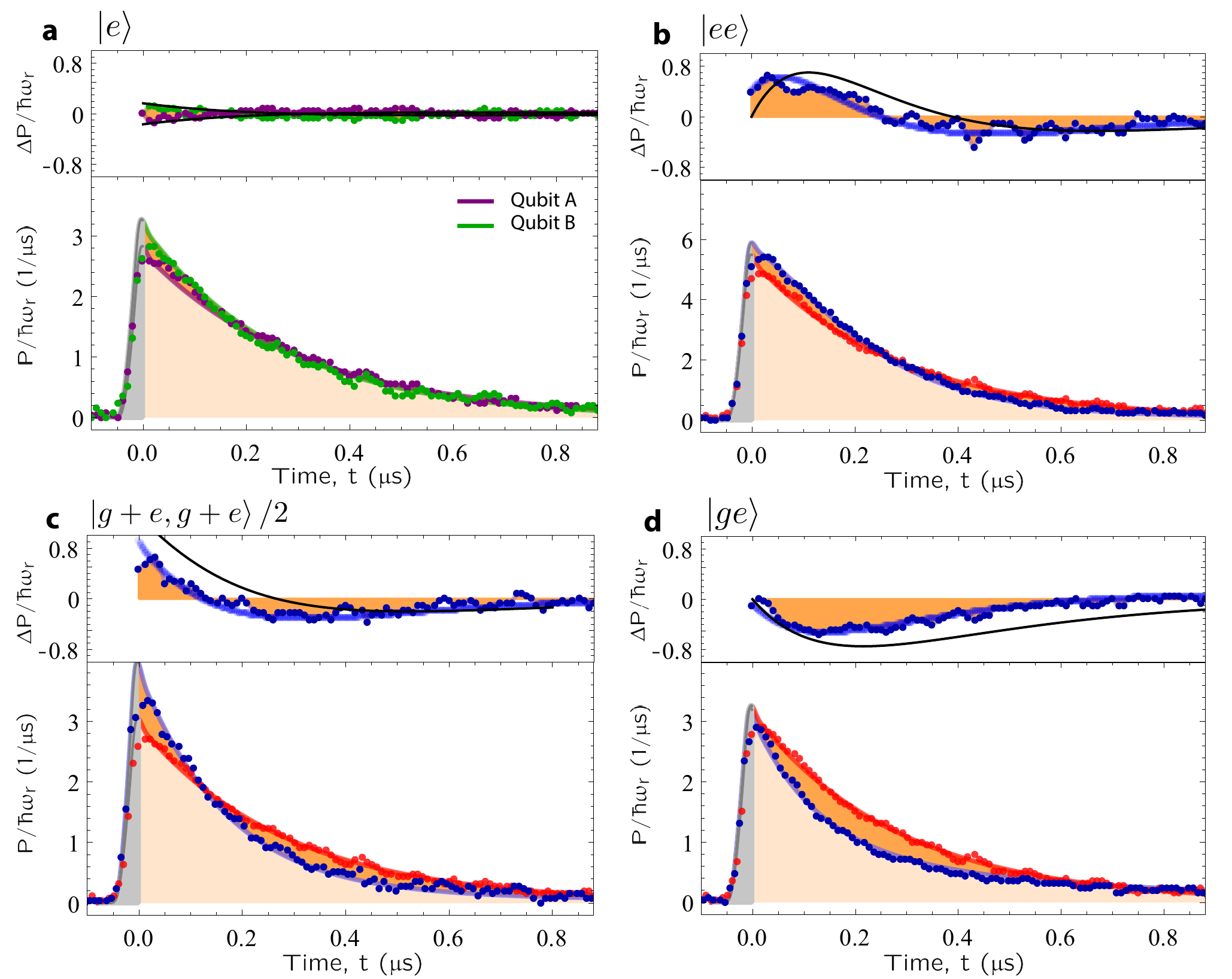}
  \caption{\textbf{Spontaneous emission and two-qubit superradiance for the indicated initial states.} In each panel the time dependence of the emitted power $P$ for a given initial state (bottom) and the deviation $\Delta P$ (top) from the average single qubit power (red points) is shown. Data (blue dots) is compared to a simple rate equation model (solid black lines) and full master equation simulations (solid blue lines), see text for details. \textbf{a}, Individual decay of qubit A (purple) or B (green) prepared in state $|e\rangle$.
  Collective decay for initial states \textbf{b} $|ee\rangle$, \textbf{c} $(|g\rangle+|e\rangle)(|g\rangle+|e\rangle)/2$, and \textbf{d} $|ge\rangle$. The orange area indicates the difference of the collective two-qubit decay with respect to the mean individual decay (dashed red line).
  For time $t<0$ (greyed-out area), the emission dynamics is governed by the initial field build-up, which is not considered in the upper parts of each panel. \chd{All data was normalized by the the same constant, extracted by matching the emitted energy of the mean individual decay to what is expected from the master equation. The theoretical curves then are scaled by $s$ to include variations in our detection efficiency, where in (b) s=0.9, in (c) s=0.94 and in (d) s=1.07. The reference curve of the mean individual decay was scaled accordingly.}
  }
  \label{fig:timetraces}
\end{figure*}

In the quickly developing field of circuit quantum electrodynamics \cite{Schoelkopf2008}, in which artificial atoms realized as superconducting qubits are coupled to microwave photons, the condition $R \sim \lambda$ or even $\ll \lambda$ is realized. Moreover experiments take advantage of the fact that emitters can be localized in a one-dimensional (1D) architecture instead of in three dimensions (3D). In particular, for 1D superconducting transmission lines single microwave photons can propagate with small loss in forward or backward direction while strong interactions can be maintained over larger distances \cite{vanLoo2013, Lalumiere2013}. As a consequence, in circuit QED experiments, super- and subradiant states have been selectively prepared in the strong coupling regime of cavity QED \cite{Filipp2011a} as well as in 1D free space \cite{vanLoo2013}. The yet largely unexplored bad (or fast) cavity limit \cite{Haroche2006}, where the cavity decay rate $\kappa$ is much larger than the coupling strength $g$ and the rates for non-radiative atomic decay $\Gamma_{\rm{nr}}$ and pure dephasing $\Gamma^*$ ($\kappa\gg g\gg\Gamma_{\rm{nr}},\Gamma^*$) extends between those two regimes.

\begin{figure}[t!]
  \centering
  \includegraphics[width=1\columnwidth]{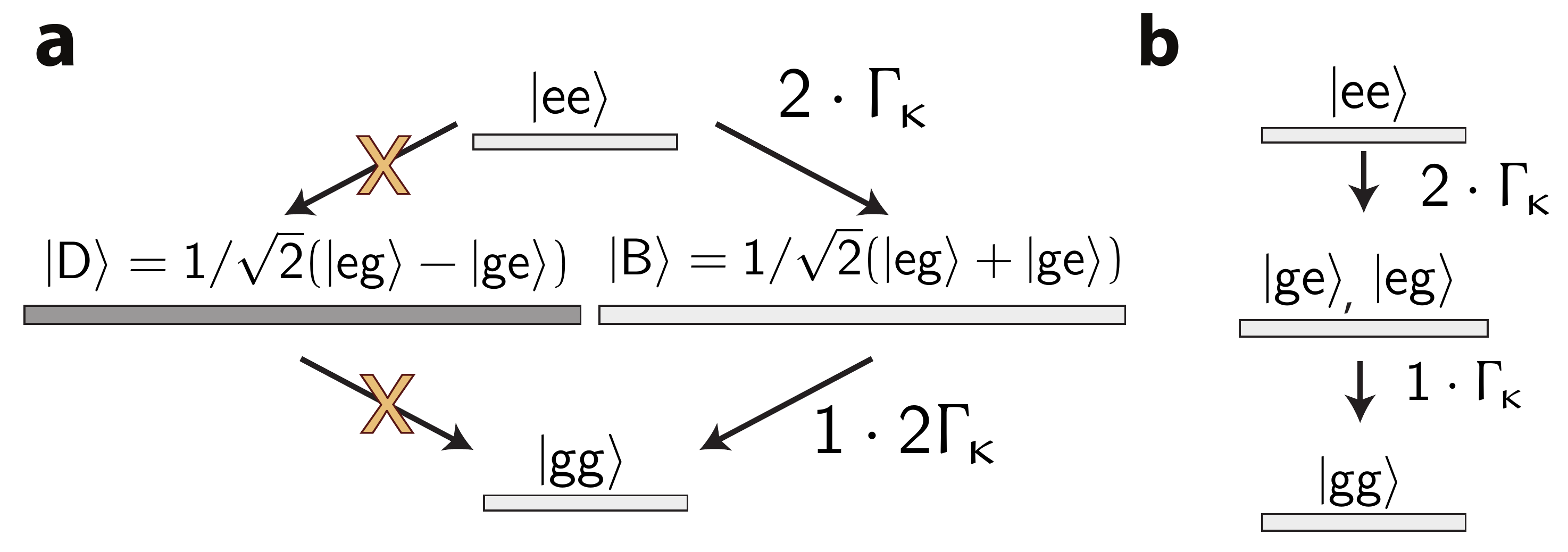}
  \caption{\textbf{Two qubit level scheme and decay channels.} (a) Eigenstates in the coupled basis. All allowed transitions happen via the bright state. The dark state does not couple to the cavity field mode.(b) Eigenstates in the uncoupled basis. The transition rates are always proportional to the number of excitations in the system.}
  \label{fig:levelscheme}
\end{figure}

We have investigated superradiance of a pair of emitters in the bad cavity limit of circuit QED as discussed theoretically \cite{Bonifacio1971,Lehmberg1970,Delanty2011} (Fig.~\ref{fig:sample}) with two transmon qubits coupled to a single coplanar waveguide resonator (see methods for details) \chd{with experimentally extracted parameters $\Gamma_{nr}/2\pi(A,B)$ $\approx$ \unit[(0.040, 0.042)]{MHz}, $\Gamma^*/2\pi(A,B)$ $\approx$ \unit[(0.25,0.27)]{MHz}, $g/2\pi(A,B)$ $\approx$ \unit[(3.5,3.7)]{MHz}, and $\kappa/2\pi$ $\approx$ \unit[43]{MHz}.}
For a first characterization of the system properties we have measured the average transmittance of the cavity with the transition frequency $\omega_{A}$ of qubit A tuned to the center frequency of the resonator $\omega_r/2\pi$ $\approx$ \unit[7.064]{GHz} while the second qubit B is kept off-resonant at $\omega_{B,0}$ $\approx$ \unit[7.41]{GHz}. The measured transmission spectrum is plotted \textsl{vs.} the frequency $\omega_p$ of a weak external probe field (Fig.~\ref{fig:transspec}). It fits well to the expected steady-state transmission function \cite{Carmichael2008Book}. The width of the broad Lorentzian peak is set by the cavity decay rate $\kappa/2 \pi$ while the narrow Lorentzian dip has a width of $w=2 \Gamma_2+4 g^2/(\kappa-2 \Gamma_2)$ with $\Gamma_2=\Gamma_{\rm{nr}}/2+\Gamma^*$ governed by the non-radiative qubit decay rate $\Gamma_{\rm{nr}}$ and pure dephasing rate $\Gamma^*$. The minimum transmission $d$ on resonance ($\omega_A=\omega_r$) is given by $\Gamma_2/(\Gamma_\kappa+\Gamma_2)$ where $\Gamma_\kappa=4g^2/\kappa$ is defined as the Purcell induced decay rate on resonance in the bad cavity limit. Physically, the distinct shape of the measured spectrum is understood in terms of atom enhanced absorption \cite{rice1988}, which is closely related to electromagnetically induced transparency \cite{Fleischhauer2005} or cavity induced transparency \cite{rice1996}. Intuitively, the coherent scattering of the probe field detuned by the same frequency but with opposite sign from the excited state doublet $(|1+\rangle,|1-\rangle)$ formed by the long-lived qubit resonantly coupled to the bad cavity (Fig.~\ref{fig:transspec}) leads to the dip in the spectrum due to destructive interference \cite{Fleischhauer2005}. \chd{It is worth noting that the spectrum can also be fully explained by the linear response of a driven resonator mode in the presence of dispersion and absorption \cite{Zhu1990} and does not necessarily require a quantum mechanical treatment.}

In a next step, we have explored the Purcell-enhanced spontaneous decay of the individual qubits. For this purpose the qubits were prepared in their excited state $|e\rangle$ by applying a $\pi$-pulse through a separate gate line (green in Fig.~\ref{fig:sample}) and tuned into resonance with the cavity by applying a magnetic field pulse using a dedicated flux line (red in Fig.~\ref{fig:sample}). In the limit $\kappa\gg g\gg \Gamma_{\rm{nr}}$, the single excited qubit shows exponential decay of the detected power $P$ (Fig.~\ref{fig:timetraces}a) with a rate of $\Gamma_\kappa=\kappa g^2/|\frac{\kappa}{2} + i \Delta_r|^2$ \cite{Delanty2011,Heinzen1987,Carmichael2008Book}. To slow down the qubit decay with respect to the bandwidth of our acquisition system we have performed the measurements at a small qubit/cavity detuning of $\Delta_r/2\pi=(\omega_{A/B}-\omega_r)/2\pi=$ \unit[25]{MHz}. The time dependence of the individual qubit decays are very similar with differences limited only by a small spread in individual coupling rates $g$. \chd{By numerically fitting the master equation simulation to the individual decays we have extracted the non-radiative decay rates $\Gamma_{nr}(\Delta=\mbox{\unit[25]{MHz.}})/2\pi$=\unit[(0.04, 0.042)]{MHz}, which are small compared to the radiative decay rates of $\Gamma_{\kappa}(\Delta=\mbox{\unit[25]{MHz.}})/2\pi$=\unit[(0.48, 0.54)]{MHz}.} The deviation of the power
$\Delta P(t)=P_0~\mathrm{e}^{-\Gamma^{\rm{A,B}}_\kappa t}-\bar{P}(t)$ emitted from the individual emitters from their mean $\bar{P}$ is plotted versus $t$ in the upper panel of Fig.~\ref{fig:timetraces}a. The normalization is given by $P_0=\hbar\omega \Gamma_\kappa$. These data sets serve as a reference for comparison with the superradiant decay of two qubits.

When both qubits are prepared in the state $|ee\rangle$ and tuned synchronously into resonance with the resonator we observe the characteristic collective superradiant decay of the two-qubit ensemble \cite{Gross1982,Delanty2011}. First we note that the emitted power level is approximately twice as large as in the single qubit case (Fig.~\ref{fig:timetraces}b) with an enhancement of the power level relative to the single qubit case at early times and a reduction at later times, which is also displayed in the upper panel of Fig.~\ref{fig:timetraces}b. In addition we note that the two-qubit collective decay begins at a rate smaller but then speeds up to values larger than the single qubit decay rate. Both features are qualitatively expected for small ensemble superradiance and are also in quantitative agreement with a master equation simulation taking into account the measured qubit relaxation and dephasing rates (blue line) and an analytic approximation
$\Delta P(t)=2P_0~\mathrm{e}^{-2\bar\Gamma_\kappa t}(1+2\bar\Gamma_\kappa t)-2 \bar{P}(t)$ (black line) \cite{Gross1982}. \chd{Intuitively, the decay process starting out at a small rate and speeding up can be understood as due to the qubits dipoles with initially undefined phase synchronizing during the decay through their interaction, which gives rise to correlations, naturally linked to the presence of entanglement, since the only allowed decay channel for the $\ket{ee}$ state is via the entangled bright state $\ket{B}=\ket{ge}+\ket{eg}$ (see figure \ref{fig:levelscheme}a). Due to the correlations the transition rate from $\ket{B}$ to $\ket{gg}$ is two times larger than the single decay rate out of the states $\ket{ge}$ or $\ket{eg}$ respectively (see figure \ref{fig:levelscheme}b). Both $\ket{B}$ and $\ket{ge}$ contain the same number of excitations. It is therefore apparent that the superradiant decay cannot follow a purely exponential dependence as the decay rate is not always proportional to the number of excitations. This intuitive argument can also be verified experimentally by initially preparing the two qubits in superposition states $(\ket{g}+\exp^{i \varphi_{A,B}}\ket{e})/\sqrt{2}$ with well defined phases. If the relative phase of the dipoles $\Delta\varphi=\varphi_A-\varphi_B$ is adjusted to $0$, the superradiant decay occurs at a single enhanced rate much earlier (Fig.~\ref{fig:timetraces}c) as the initial state, written in the coupled basis, already contains a $\ket{B}$ state part and thus is provided with correlations right away.} Also in this case the observed decay dynamics are in good agreement with theory
$\Delta P(t)=P_0~\mathrm{e}^{-2\bar\Gamma_\kappa t}(\tfrac{3}{2}+\bar\Gamma_\kappa t)-\bar{P}(t)$ (black lines) and master equation simulations (blue lines).

\begin{figure*}[t!]
  \centering
  \includegraphics[width=0.8\textwidth]{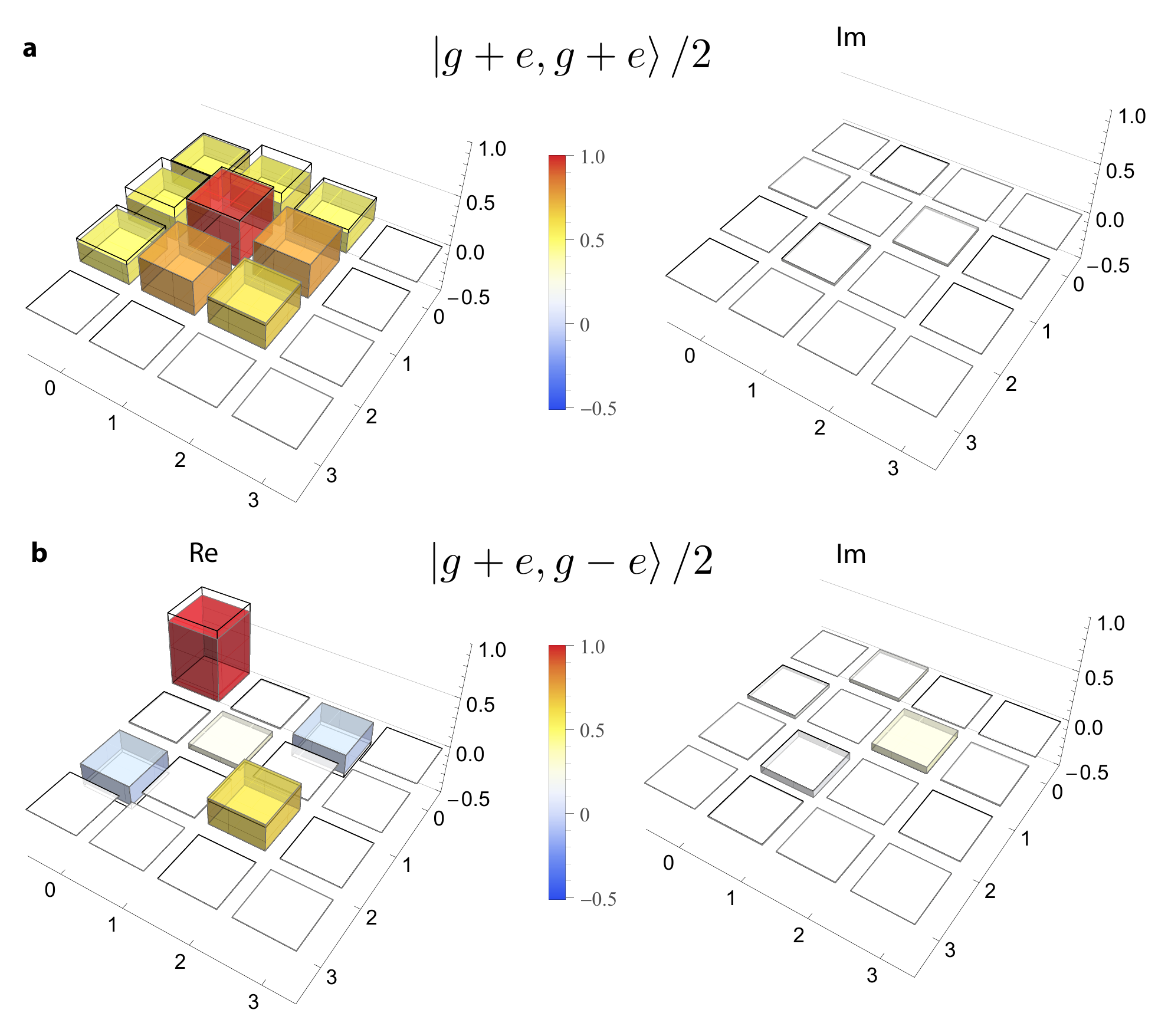}
  \caption{\textbf{Density matrix of the emitted field.} \chd{ Real and imaginary part of the measured density matrix $\rho$ (colored bars) compared to expected values $\rho_+$ and $\rho_-$(wire frames) for initial qubit states \textbf{a} $(|g\rangle+|e\rangle)(|g\rangle+|e\rangle)/2$ and \textbf{b} $(|g\rangle+|e\rangle)(|g\rangle-|e\rangle)/2$.}}
  \label{fig:statetomo}
\end{figure*}

Notably, the physical system investigated here also allows for the experimental investigation of a situation which Dicke has denoted as single atom superradiance in his initial Gedankenexperiment. In an effect surprising at the time a single emitter in the excited state $|e\rangle$ is predicted to decay at an enhanced rate in the presence of a second emitter, even when that second emitter is in its ground state $|g\rangle$ (Fig.~\ref{fig:timetraces}d). Here the initial state $|ge\rangle$ can be decomposed into a superposition $(\ket{B}+\ket{D})/\sqrt{2}$ of a bright $\ket{B}=(\ket{ge}+\ket{eg})/\sqrt{2}$ and a dark state $\ket{D}=(\ket{ge}-\ket{eg})/\sqrt{2}$.  Half of the initial excitation remains trapped in $\ket{D}$ while the other half decays at twice the rate from the state $\ket{B}$, as pointed out by Dicke in his original argument \cite{Dicke1954}. Again the measured data is in agreement with solutions of simple coupled rate equations, namely $\Delta P(t)=P_0~\mathrm{e}^{-2\bar\Gamma_\kappa t}-\bar{P}(t)$ (Fig.~\ref{fig:timetraces}, black lines) \cite{Gross1982}.
\chd{Ideally one would expect to measure 50\% of the initial excitation to be trapped. In our experiment we measure the emitted energy to be \unit[0.707]{photons}. This value is consistent with the master equation expectation of \unit[0.709]{photons} and the deviation from the ideal case is attributed to the finite dephasing rate $\Gamma^*$, which leads to a lifting of the dark state and a resulting leakage of the excitation into the cavity mode.}

To further characterize the superradiant decay of the two-qubit ensemble, we have fully reconstructed the single mode density matrix of the emitted field using a statistical analysis of the measured quadrature amplitudes \cite{Eichler2012}. Any initial pure and separable two qubit state brought into resonance with the cavity can be expressed in the coupled atomic basis states as $\alpha\ket{gg}+\delta\ket{D}+\beta\ket{B}+\gamma\ket{ee}$. The resulting photonic state is in general a mixed state obtained by tracing over $\ket{D}$ and reads $\delta^2 \ket{0}\bra{0} +(1-\delta^2)\ket{\Psi_B}\bra{\Psi_B}$ with $\Psi_B=(1-\delta^2)^{-1/2}(\alpha\ket{0}+\beta\ket{1}+\gamma\ket{2})$. We have reconstructed the density matrix of such output states for both qubits initially in the state $(\ket{g}+\ket{e})/\sqrt{2}$ (Fig.~\ref{fig:statetomo}a). The reconstructed density matrix clearly shows that the emitted field consists of zero, one, and two photon Fock states and features pronounced coherences between those states (colored bars) in good agreement with the expected output state $\rho_+=(1/2\ket{0}+1/\sqrt{2}\ket{1}+1/2\ket{2})(1/2\bra{2}+1/\sqrt{2}\bra{1}+1/2\bra{2})$ (wireframe).
The state fidelity of the measured state $\rho$ with respect to $\rho_+$ is $F=(Tr\sqrt{\sqrt{\rho_+}\rho\sqrt{\rho_+}})^2=0.94$.

 Initially preparing the two qubits in equal superposition states $(\ket{g}+\ket{e})/\sqrt{2}\otimes(\ket{g}-\ket{e})/\sqrt{2}$ out of phase by $\pi$, either zero or two photons are emitted, displaying a coherent component as well, while the probability for measuring a single photon vanishes (Fig.~\ref{fig:statetomo}b), compatible with the expected mixed state of the form $\rho_-=1/2\ket{0}\bra{0}+1/4(\ket{0}-\ket{2})(\bra{2}-\bra{0})$ with a fidelity of F=0.94.

The experimental observation of superradiance for a microscopic two-qubit ensemble prepared in a set of well defined initial states demonstrated here represents a close to ideal realization of Dicke's pioneering ideas. The control of superradiance of small ensembles may prove essential for experiments exploring entanglement via dissipation \cite{harkonen2009,gonzalez2013}, measurement induced entanglement \cite{schneider2002,Julsgaard2012b}, teleportation via superradiance \cite{chen2005}, two-color superradiance \cite{hayn2011} or time-resolved correlations \cite{temnov2009}.

\section{Methods summary}
To perform the presented experiments a circuit QED sample with two qubits of the transmon type \cite{Koch2007} interacting with a coplanar waveguide resonator was fabricated in two-dimensional geometry using standard techniques. The resonator is weakly coupled to an input and overcoupled to an output line resulting in large decay rate. A smaller than typical coupling rate $g$ was realized by creating a qubit geometry in which island and reservoir couple almost identically to the resonator. The qubits were positioned at field maxima of the first harmonic mode of the resonator.

The pulse scheme used for individual control and read out of the qubits is similar to the one used for observing collective dynamics in strong coupling circuit QED \cite{Mlynek2012}. In their idle position, both qubits are tuned to their maximum transition frequencies of $\omega_{A,0}\approx$ \unit[8.20]{GHz} and $\omega_{B,0}\approx$ \unit[7.40]{GHz} by using miniature superconducting coils mounted on the backside of the chip which allow for individual flux biasing of the qubit SQUID loops. Single-qubit operations are realized using \unit[12]{ns} long resonant microwave pulses. Qubit transition frequencies were tuned on ns timescales by injecting current pulses into the flux gate line (Fig.~\ref{fig:sample}).

If the flux pulse is chosen to tune the qubit to $\omega_r$ the qubit state $\alpha\ket{g}+\beta\ket{e}$ is transformed into a purely photonic state $\alpha\ket{0}+\beta\ket{1}$. This mapping is justified because in the bad
cavity limit the resonator mode can be adiabatically eliminated and thus treated as a simple decay channel.


\chd{To measure the field we used a heterodyne setup which extracts the complex amplitude, consisting of the experimental signal including the amplifier noise dominated by a high-electron-mobility transistor amplifier with gain \unit[30]{dB} at \unit[4]{K}. Additional amplification by \unit[60]{dB} is performed at room temperature. The microwave signal is then mixed down to \unit[25]{MHz}, again amplified by \unit[30]{dB}, digitized using an analogue to digital converter with a time resolution of \unit[10]{ns} and finally processed with field programmable gate array (FPGA) electronics, which also digitally converts the signal down to DC and uses a 4 point square filter to eliminate frequency components higher than \unit[25]{MHz}. To extract the photon number the FPGA calculates the square of the complex amplitude in real time and then averages over multiple instances of the same experiment assuming that the noise and the signal are fully uncorrelated. The noise floor is determined from an off-measurement where no photons are generated and can be numerically accounted for.}

The tomographic measurements were performed using the techniques discussed in Ref.~\cite{Eichler2012} making use of a parametric amplifier \cite{Eichler2014} operated in the phase preserving mode to reduce the required integration time.

\section{Acknowledgments}
\begin{acknowledgments}
We would like to acknowledge financial support by ETH Zurich and thank A. Imamoglu, G.Milburn and J. Twamley for discussion and valuable comments.
\end{acknowledgments}

\bibliographystyle{naturemagNOURL}


\end{document}